\documentclass[italian,english]{article}
\usepackage[latin9]{inputenc}
\usepackage{color}
\usepackage{amssymb}

\makeatletter
\newcommand{\lyxaddress}[1]{
\par {\raggedright #1
\vspace{1.4em}
\noindent\par}
}

\makeatother

\usepackage{babel}

\begin{document}

\title{\textbf{Primordial gravity's breath}}

\author{\textbf{Christian Corda}%
\thanks{partially supported by a Research Grant of The R. M. Santilli Foundation
Number RMS-TH-5735A2310%
}}

\maketitle

\lyxaddress{\begin{center}
International Institute for Theoretical Physics and Advanced Mathematics
Einstein-Galilei, Via Santa Gonda, 14 - 59100 PRATO, Italy
\par\end{center}}

\begin{center}
\textit{E-mail address:} \textcolor{blue}{cordac.galilei@gmail.com}
\par\end{center}
\begin{abstract}
In a recent paper the Laser Interferometer Gravitational-wave Observatory
(LIGO) Scientific Collaboration (LSC) obtained an upper limit on the
stochastic gravitational-wave background (SGWB) of cosmological origin
by using the data from a two-year science run of the LIGO. Such an
upper limit rules out some models of early Universe evolution, like
the ones with relatively large equation-of-state parameter and the
cosmic (super) string models with relatively small string tension
arising from some String Theory's models. This was also an upper limit
for the SGWB which is proposed by the Pre-Big-Bang Theory.

Another upper bound on the SGWB which is proposed by the Standard
Inflationary Model is well known and often updated by using the Wilkinson
Microwave Anisotropy Probe (WMAP) data.

By using a conformal treatment, which represents a variation of early
works, we release a formula that directly connects the average amplitude
of the SGWB with the Inflaton field in the Standard Inflationary Scenario
of General Relativity and an external Inflaton field. Then, by joining
this formula with the equation for the characteristic amplitude $h_{c}$
for the SGWB, the upper bounds on the SGWB from the WMAP and LSC data
will be translated in lower bounds on the Inflaton field.

The results show that the value of the Inflaton field that arises
from the WMAP bound on the SGWB is totally consistent with the famous
\textit{slow roll} condition on Inflation. On the other hand, the
value of the Inflaton field that arises from the LSC bound on the
SGWB could be not consistent with such a condition. 

In any case, the analysis in this paper shows that the detection of
the SGWB will permit a direct measure of the value of the Inflaton
field by giving an extraordinary precious and precise information
about the early Universe's dynamics. In other words, the detection
of the SGWB will permit \emph{to auscultate} \emph{the primordial
gravity's breath}.
\end{abstract}

\section{Introduction}

The scientific community aims in a first direct detection of gravitational
waves (GWs) in next years (for the current status of GWs interferometers
see \cite{key-1}) confirming the indirect, Nobel Prize Winner, proof
of Hulse and Taylor \cite{key-2}.

Detectors for GWs will be important for a better knowledge of the
Universe and either to confirm or to rule out, in an ultimate way,
the physical consistency of General Relativity, eventually becoming
an observable endorsement of Extended Theories of Gravity \cite{key-3}.

It is well known that an important potential source of gravitational
radiation is the relic SGWB \cite{key-4}. The potential existence
of such a relic SGWB arises from general assumptions that mix principles
of classical gravity with principles of quantum field theory \foreignlanguage{italian}{\cite{key-5,key-6,key-7}}.
As the zero-point quantum oscillations, which produce the relic SGWB,
are generated by strong variations of the gravitational field in the
early Universe, the potential detection of this relic SGWB is the
only way to learn about the evolution of the primordial Universe,
up to the bounds of the Planck epoch and the initial singularity \cite{key-4,key-7}.
In fact, this kind of information is inaccessible to standard astrophysical
observations \cite{key-4,key-7,key-8}. The importance of this relic
signal in cosmological scenarios has been discussed in an elegant
way in \cite{key-8}.

The inflationary scenario for the early Univers\foreignlanguage{italian}{e
\cite{key-9,key-10}}, which is tuned in a good way with the WMAP
data on the Cosmic Microwave Background Radiation (CMBR) (in particular
exponential inflation and spectral index $\approx1$ \foreignlanguage{italian}{\cite{key-11}),}
amplified the zero-point quantum oscillations and generated the relic
SGWB \cite{key-6,key-7}.

A recent paper, which has been written by the LSC \cite{key-4}, has
shown an upper limit on the SGWB by using the data from a two-year
science run of LIGO. Such an upper limit rules out some models of
early Universe evolution, like the ones with relatively large equation-of-state
parameter and the cosmic (super) string models with relatively small
string tension arising from some string theory models. It results
also an upper limit for the SGWB which is proposed by the Pre-Big-Bang
Theory (see \cite{key-4} for details).

Another well known upper bound on the SGWB arises from the Standard
Inflationary Model. Such an upper bound is often updated by using
the WMAP data \cite{key-4,key-12}. 

In this paper a formula that directly connects the average amplitude
of the SGWB with the Inflaton field will be obtained by using a variation
of a conformal treatment analysed in \cite{key-13} and \cite{key-14}.
By using such a formula and the equation for the characteristic amplitude
$h_{c}$ for the SGWB \cite{key-15}, the upper bounds on the SGWB
from the WMAP and LSC data will be translated in lower bounds on the
Inflaton field.

Our results show that the value of the Inflaton field that arises
from the WMAP bound on the SGWB is totally consistent with the famous
\textit{slow roll} condition on Inflation \foreignlanguage{italian}{\cite{key-9,key-10}},
while the value of the Inflaton field that arises from the LSC bound
on the SGWB could not be consistent with this condition.

The analysis in this paper shows that the detection of the SGWB will
permit, ultimately, a direct measure of the value of the Inflaton
field by giving an extraordinary precious and precise information
on the early Universe's dynamics. In other words, the detection of
the SGWB will permit \emph{to auscultate} \emph{the primordial gravity's
breath.}

\section{The spectrum and the conformal treatment}

Considering a relic SGWB, it can be characterized by a dimensionless
spectrum \cite{key-4,key-7,key-8}. The more recent values for the
spectrum that arises from the WMAP data can be found in refs. \cite{key-4,key-12}.
In such papers it is (for a sake of simplicity, in this paper natural
units are used, i.e. $8\pi G=1$, $c=1$ and $\hbar=1$) \begin{equation}
\Omega_{gw}(f)\equiv\frac{1}{\rho_{c}}\frac{d\rho_{gw}}{d\ln f}\leq10^{-13}\label{eq: spettro}\end{equation}

where \begin{equation}
\rho_{c}\equiv3H_{0}^{2}\label{eq: densita critica}\end{equation}

is the (actual) critical density energy, $\rho_{c}$ of the Universe,
$H_{0}$ the actual value of the Hubble expansion rate and $d\rho_{gw}$
the energy density of relic GWs in the frequency range $f$ to $f+df$.
This is the upper bound on the SGWB that observations put on the Standard
Inflationary Model, i.e exponential Inflation and spectral index $\approx1$.

An higher bound results from the LIGO Scientific Community data in
ref. \cite{key-4}:

\begin{equation}
\Omega_{gw}\leq6.9*10^{-6}.\label{eq: spettro LIGO}\end{equation}

This bound is at $95\%$ confidence in the frequency band $41.5-169.25Hz$,
(see \cite{key-4} for details).

In this case, the value is an upper limit for the SGWB which arises
from the Pre-Big-Bang Theory \cite{key-4,key-16}. It also rules out
some models of early Universe evolution, like the ones with relatively
large equation of state parameter and the cosmic (super) string models
with relatively small string tension arising from some string theory
models (see \cite{key-4}and references within). 

We will consider a variation of a computation in \cite{key-13}. In
such a paper a conformal treatment has been applied to Scalar Tensor
Gravity. A similar computation was also performed in \cite{key-14}
in the framework of $f(R)$ Theories of Gravity.

However, Scalar Tensor Gravity and $f(R)$ Theories are only particular
cases where an external scalar field works like Inflaton, other cases
could be, for example, the Higgs potential and a self-interacting
scalar field. 

In this work we discuss the Standard Model's case, in which the scalar
field (Inflaton) arises from field theory \foreignlanguage{italian}{\cite{key-9}}. 

In the Standard Scenario inflation can of solve many of the initial
value, or \textquoteleft{}fine-tuning\textquoteright{}, problems of
the hot Big Bang model \foreignlanguage{italian}{\cite{key-9}}. The
fundamental assumption is that there is some mechanism to bring about
the negative pressure state needed for quasi-exponential growth of
the scale factor \foreignlanguage{italian}{\cite{key-9}}. Inflaton
is the name given to a relic scalar field $\varphi$, since its origin
does not have to originate with a specified particle theory \foreignlanguage{italian}{\cite{key-9}}.
The original hope was that $\varphi$ would help to determine the
correct particle physics models but current model building does not
necessarily require specific particle phenomenology \foreignlanguage{italian}{\cite{key-9}}.
This is actually an advantage for the Standard Inflationary Scenario,
as it retains its power to solve the initial value problems, yet it
could arise from any arbitrary source (i.e., any arbitrary Inflaton)
\foreignlanguage{italian}{\cite{key-9}}. In field theory, we consider
a Lagrangian density, as opposed to the usual Lagrangian from classical
mechanics \foreignlanguage{italian}{\cite{key-9}}. In fact, in field
theory scalar fields are taken to be continuous fields, whereas the
Lagrangian in mechanics is usually based on discrete particle systems.
The Lagrangian $L$ is related to the Lagrangian density $\mathcal{L}$
by \foreignlanguage{italian}{\cite{key-9}}

\selectlanguage{italian}%
\begin{equation}
L=\int\mathcal{L}d^{3}x.\label{eq: lagrangian}\end{equation}

The scalar field is represented by a continuous function \foreignlanguage{english}{$\varphi(x,t)$}
which can be real or complex. Given a potential density of the field
$V(\varphi)$, \foreignlanguage{english}{$\mathcal{L}$} reads \cite{key-9}

\selectlanguage{english}%
\begin{equation}
\mathcal{L}=\frac{1}{2}\partial_{\mu}\varphi\partial^{\mu}\varphi-V(\varphi).\label{eq: scalar lagrangian}\end{equation}
\foreignlanguage{italian}{ }

Let us consider the standard Einstein-Hilbert action of General Relativity
which is \cite{key-17,key-18,key-19}

\begin{equation}
S=\int d^{4}x\sqrt{-g}(R+\mathcal{L}_{m}),\label{eq: standard}\end{equation}

where $\mathcal{L}_{m}$ is the Lagrangian of the matter.

One can define the {}``conformal scalar field'' like a logarithm
of the Inflaton field \cite{key-13} \begin{equation}
e^{2\Phi}\equiv\varphi.\label{eq: rescaling}\end{equation}

By applying the conformal transformation \cite{key-19}

\begin{equation}
\tilde{g}_{\alpha\beta}=e^{2\Phi}g_{\alpha\beta}\label{eq: conforme}\end{equation}
to the action (\ref{eq: standard}) the conformal equivalent Hilbert-Einstein
action \cite{key-19} \begin{equation}
A=\int\frac{1}{2k}d^{4}x\sqrt{-\widetilde{g}}[\widetilde{R}+L_{1}(\Phi,\Phi_{;\alpha})],\label{eq: conform}\end{equation}

is obtained. In this way, the analysis can be translated in a conformal
frame. $L_{1}(\Phi,\Phi_{;\alpha})$ is the conformal scalar field
contribution derived from

\begin{equation}
\tilde{R}_{\alpha\beta}=R_{\alpha\beta}+2(\Phi_{;\alpha}\Phi_{;\beta}-g_{\alpha\beta}\Phi_{;\delta}\Phi^{;\delta}-\frac{1}{2}g_{\alpha\beta}\Phi^{;\delta}{}_{;\delta})\label{eq: conformRicci}\end{equation}

and \begin{equation}
\tilde{R}=e^{-2\Phi}(R-6\square\Phi-6\Phi_{;\delta}\Phi^{;\delta}).\label{eq: conformRicciScalar}\end{equation}
In the re-scaled action (\ref{eq: conform}) the matter contributions
have not been considered because our interaction with GWs concerns
the linearized theory in vacuum. 

Following \cite{key-13}, it is well known that the gravity-wave amplitude
$h_{i}^{j}$ (in the following we will consider the {}``plus'' polarization
$h_{+}$) is a conformal invariant and that the d'Alembert operator
transforms as \cite{key-13,key-14}

\begin{equation}
\widetilde{\square}=e^{-2\Phi}(\square+2\Phi^{;\alpha}\partial_{;\alpha}).\label{eq: quadratello}\end{equation}

Thus, the background changes in the conformal frame while the tensor
wave amplitude is fixed. 

In order to study the cosmological stochastic background, the operator
(\ref{eq: quadratello}) has to be specified for a Friedman-Robertson-Walker
metric \cite{key-13,key-14}, obtaining

\begin{equation}
\ddot{h}_{+}+(3H+2\dot{\Phi})\dot{h}_{+}+k^{2}a^{-2}h_{+}=0,\label{eq: evoluzione h}\end{equation}

being $\square=\frac{\partial}{\partial t^{2}}+3H\frac{\partial}{\partial t}$,
$a(t)$ the scale factor and $k$ the wave number. 

Considering the conformal time $d\eta=dt/a$, Eq. (\ref{eq: evoluzione h})
reads\begin{equation}
\frac{d^{2}}{d\eta^{2}}h_{+}+2\frac{\gamma'}{\gamma}\frac{d}{d\eta}h_{+}+k^{2}h_{+}=0,\label{eq: evoluzione h 3}\end{equation}

where $\gamma=ae^{\Phi}$. Inflation implies $\eta=\int(dt/a)=1/(aH)$
and $\frac{\gamma'}{\gamma}=-\frac{1}{\eta}$ \cite{key-13,key-14}. 

Eq. (\ref{eq: evoluzione h 3}) is formally equal to the equation
of a damped harmonic oscillator $\mu(t)$ \begin{equation}
\ddot{\mu}+K\dot{\mu}+\omega_{0}^{2}\mu=0,\label{eq: oscillatore}\end{equation}

where $K$ is the damping constant and $\omega_{0}$ the proper frequency
of the harmonic oscillator, but in the case of Eq. (\ref{eq: evoluzione h 3})
the \emph{effective damping constant} $2\frac{\gamma'}{\gamma}$ depends
on the conformal time. Hence, we are working with an \emph{effective
damped harmonic oscillator}. In any case, the solution of Eq. (\ref{eq: evoluzione h 3})
has been found in \cite{key-13,key-14} \begin{equation}
h_{+}(\eta)=k{}^{-3/2}\sqrt{2/k}[C_{1}(\sin k\eta-\cos k\eta)+C_{2}(\sin k\eta+\cos k\eta)].\label{eq: sol ev h2}\end{equation}

Inside the $1/H$ radius it is $k\eta\gg1.$ Furthermore, considering
the absence of GWs in the initial vacuum state, only negative-frequency
modes are present and then the adiabatic behavior is \cite{key-13,key-14}
\begin{equation}
h_{+}=k{}^{1/2}\sqrt{2/\pi}\frac{1}{aH}C\exp(-ik\eta).\label{eq: sol ev h3}\end{equation}

At the first horizon crossing ($aH=k$ at $t=10^{-22}$ second after
the Initial Singularity, \cite{key-7}), the averaged amplitude $A_{h_{+}}=(k/2\pi)^{3/2}h_{+}$
of the perturbations is \begin{equation}
A_{h_{+}}=\frac{1}{2\pi{}^{2}}C\label{eq: Ah-1}\end{equation}

when the scale $a/k$ grows larger than the Hubble radius $1/H,$
the growing mode of evolution is constant ({}``frozen'', see \cite{key-13,key-14}).
This situation corresponds to the limit $-k\eta\ll1$ in equation
(\ref{eq: sol ev h2}). 

The amplitude $A_{h_{+}}$ of the wave is preserved until the second
horizon crossing after which it can be observed, in principle, as
an anisotropy perturbation of the CMBR \cite{key-7,key-8}. It can
be shown that $\frac{\delta T}{T}\leq A_{h_{+}}$ is an upper limit
to $A_{h_{+}}$ since other effects can contribute to the background
anisotropy \cite{key-13,key-14}. Then, it is clear that the only
relevant quantity is the initial amplitude $C$ in equation (\ref{eq: sol ev h3})
which is conserved until the re-enter. Such an amplitude directly
depends on the fundamental mechanism generating perturbations that
depends on the Inflaton scalar field which generates inflation.

Considering a single monocromatic GW, its zero-point amplitude is
derived through the commutation relations \cite{key-13,key-14} \begin{equation}
[h_{+}(t,x),\pi_{h_{+}}(t,y)]=i\delta^{3}(x-y)\label{eq: commutare}\end{equation}

calculated at a fixed time $t.$ 

As it is \cite{key-13,key-14}

\begin{equation}
\pi_{h_{+}}=e^{2\Phi}a^{3}\dot{h}_{+},\label{eq: pi h}\end{equation}

equation (\ref{eq: commutare}) reads\begin{equation}
[h_{+}(t,x),\dot{h}_{+}(y,y)]=i\frac{\delta^{3}(x-y)}{e^{2\Phi}a^{3}}\label{eq: commutare 2}\end{equation}

and the fields $h_{+}$ and $\dot{h}_{+}$ can be expanded in terms
of creation and annihilation operators \cite{key-13,key-14}

\begin{equation}
h_{+}(t,x)=\frac{1}{(2\pi)^{3/2}}\int d^{3}k[h_{+}(t)e^{-ikx}+h_{+}^{*}(t)e^{ikx}]\label{eq: crea}\end{equation}
\begin{equation}
\dot{h}_{+}(t,x)=\frac{1}{(2\pi)^{3/2}}\int d^{3}k[\dot{h}_{+}(t)e^{-ikx}+\dot{h}_{+}^{*}(t)e^{ikx}].\label{eq: distruggi}\end{equation}

The commutation relations in conformal time are then \cite{key-13,key-14}
\begin{equation}
[h_{+}\frac{d}{d\eta}h_{+}^{*},-h_{+}^{*}\frac{d}{d\eta}h_{+}]=i\frac{8\pi^{3}}{e^{2\Phi}a^{3}}.\label{eq: commutare 3}\end{equation}

Inserting (\ref{eq: sol ev h3}) and (\ref{eq: Ah-1}), it is $C=\sqrt{2}\pi{}^{2}He^{-\Phi}$
where $H$ and $\Phi$ are calculated at the first horizon crossing
and then \begin{equation}
A_{h_{+}}=\frac{\sqrt{2}}{2}He^{-\Phi},\label{eq: Ah2}\end{equation}

which means that the amplitude of GWs produced during inflation directly
depends on the Inflaton field being $\Phi=\frac{1}{2}\ln\varphi$
\cite{key-13}. Explicitly, it is \begin{equation}
A_{h_{+}}=\frac{H}{\sqrt{2\varphi}}.\label{eq: Ah3}\end{equation}

Thus, one immediately obtains \begin{equation}
\varphi=\frac{H^{2}}{2A_{h_{+}}^{2}}.\label{eq: fi}\end{equation}
that links directly the amplitude of relic GWs with the Inflaton scalar
field $\varphi$ which generates inflation. Then, we have re-obtained
the important Eq. (\ref{eq: fi}) in the general Standard Inflationary
context of General Relativity plus an external scalar field which
generates inflation. In this way, we have also completed the analyses
of \cite{key-13,key-14}, which concerned the particular cases of
Scalar Tensor Gravity and $f(R)$ Theories.

\section{Bounds from observations}

The equation for the characteristic amplitude $h_{c}$ is (see Equation
65 in \cite{key-15})

\begin{equation}
h_{c}(f)\simeq1.26*10^{-18}\left(\frac{1{\rm Hz}}{f}\right)\sqrt{h_{100}^{2}\Omega_{gw}(f)},\label{eq: legame ampiezza-spettro}\end{equation}

where $h_{100}\simeq0.71$ is the best-fit value on the Hubble constant
\cite{key-11}. This equation gives a value of the amplitude of the
relic SGWB in function of the spectrum in the frequency range of ground
based detectors \cite{key-15}. Such an amplitude is also the averaged
strain applied on the detector's arms by the relic SGWB \cite{key-15}.
Such a range is given by the interval $10Hz\leq f\leq10KHz$ \cite{key-1}.

Defining the average value of $h_{c}(f)$ like

\begin{equation}
A_{h_{c}}\equiv\frac{\int1.26*10^{-18}\sqrt{h_{100}^{2}\Omega_{gw}(f)}f^{-1}df}{\int df}\label{eq:average}\end{equation}

one can assume that it is $A_{h_{c}}\simeq A_{h_{+}}$ \cite{key-13}. 

In this way, it is also 

\begin{equation}
\varphi\simeq\frac{H^{2}}{2A_{h_{c}}^{2}}.\label{eq: fi 2}\end{equation}

Now, by using Eq. (\ref{eq: fi 2}), we can use the bounds (\ref{eq: spettro})
and (\ref{eq: spettro LIGO}) on the relic SGWB in order to obtain
bounds on the Inflaton field $\varphi.$ First of all, we emphasize
that a redshift correction is needed because $H$ in Eq. (\ref{eq: fi 2})
is computed at the time of the first horizon crossing, while the value
of $A_{h_{c}}$ from the WMAP and LSC data is computed at the present
time of the cosmological Era. The redshift correction on the spectrum
is well known \cite{key-7}:

\begin{equation}
\Omega_{gw}(f)=\Omega_{gw}^{0}(f)(1+z_{eq})^{-1},\label{eq: correzione}\end{equation}

where $\Omega_{gw}^{0}(f)$ is the value of the spectrum at the first
horizon crossing and $z_{eq}\simeq3200$ \cite{key-11} is the redshift
of the Universe when the matter and radiation energy density were
equal, see \cite{key-7} for details. 

Then, Eq. (\ref{eq: fi 2}) becomes

\begin{equation}
\varphi\simeq\frac{H^{2}}{2A_{h_{c}}^{2}(1+z_{eq})}.\label{eq: fi 3}\end{equation}

By considering the WMAP bound (\ref{eq: spettro}), the integrals
in Eq. (\ref{eq:average}) have to be computed in the frequency range
of ground based detectors which is the interval $10Hz\leq f\leq10KHz$.
One gets $A_{h_{c}}^{2}\simeq10^{-51}$.

By restoring ordinary units and recalling that $H\simeq10^{22}Hz$
at the first horizon crossing \cite{key-7}, at the end, from Eq.
(\ref{eq: fi 3}), we get \begin{equation}
\varphi\geq10^{2}grams.\label{eq: inflaton value}\end{equation}

This result represents a lower bound for the value of the Inflaton
field that arises from the WMAP data on the relic SGWB in the case
of Standard Inflation \cite{key-4,key-12}. 

Now, let us consider the LSC bound (\ref{eq: spettro LIGO}). Such
a bound is at $95\%$ confidence in the frequency band $41.5-169.25Hz$
\cite{key-4}, thus, in principle, we could not extend the integrals
in Eq. (\ref{eq:average}) to the total interval $10Hz\leq f\leq10KHz$.
However, it is well known that for frequencies that are smaller than
some hertz the spectrum which arises from the Pre-Big-Bang Theory
rapidly falls, while at higher frequencies the spectrum is almost
flat with a small decreasing \cite{key-4,key-16}. Thus, the integration
of Eq. (\ref{eq:average}) in the interval $10Hz\leq f\leq10KHz$
gives a solid upper bound for $A_{h_{c}}$ in these models. One gets
$A_{h_{c}}^{2}\simeq10^{-44}$. In this case, by restoring ordinary
units and putting the value $H\simeq10^{22}Hz$ in eq. (\ref{eq: fi 3})
it is \begin{equation}
\varphi\geq10^{-5}grams.\label{eq: inflaton value 2}\end{equation}

This result represents a lower bound for the value of the Inflaton
field that arises from the LSC data on the relic SGWB and it has to
be applied to the case of the Pre-Big-Bang Theory \cite{key-4,key-16}. 

It is well known that the requirement for inflation, which is $p=-\rho$
\foreignlanguage{italian}{\cite{key-9,key-10}}, can be approximately
met if one requires $\dot{\varphi}<<V(\varphi)$, where $V(\varphi)$
is the potential density of the field in Eq. (\ref{eq: scalar lagrangian}).
This leads to the famous \textit{slow-roll approximation} (SRA), which
provides a natural condition for inflation to occur \foreignlanguage{italian}{\cite{key-9,key-10}}.
The constraint on $\dot{\varphi}$ is assured by requiring $\ddot{\varphi}$
to be negligible. With such a requirement, the slow-roll parameters
are defined (in natural units) by \foreignlanguage{italian}{\cite{key-9,key-10}}

\begin{equation}
\begin{array}{c}
\epsilon(\varphi)\equiv\frac{1}{2}(\frac{V'(\varphi)}{V(\varphi)})^{2}\\
\\\eta(\varphi)\equiv\frac{V''(\varphi)}{V(\varphi)}.\end{array}\label{eq: slow-roll}\end{equation}

Then, the SRA requirements are\foreignlanguage{italian}{ \cite{key-9,key-10}}:

\begin{equation}
\begin{array}{c}
\epsilon\ll1\\
\\|\eta|\ll1,\end{array}\label{eq: slow-roll2}\end{equation}

that are satisfied when it is \foreignlanguage{italian}{\cite{key-9,key-10}}

\begin{equation}
\varphi\gg M_{Planck},\label{eq: planckiano}\end{equation}

where the Planck mass, which is $M_{Planck}\simeq2.177*10^{-5}grams$
in ordinary units and $M_{Planck}=1$ in natural units has been introduced\foreignlanguage{italian}{
\cite{key-9,key-10}}.

Then, one sees immediately that the value of the Inflaton field of
Eq. (\ref{eq: inflaton value}), that arises from the WMAP bound on
the relic SGWB, is totally in agreement with the slow roll condition
on Inflation. On the other hand, the value of the Inflaton field of
Eq. (\ref{eq: inflaton value 2}), that arises from the LSC bound
on the relic SGWB, is of the order of the Planck mass, thus, it could
not be in agreement with the slow roll condition on Inflation.

The fact that the spectrum of the relic SGWB decreases with increasing
Inflaton field is not surprising. In fact, even if the amplification
of zero-point quantum oscillations increases the spatial dimensions
of perturbations, it is well known that the curvature of the Universe
is {}``redshifted'' by Inflation, i.e. the inflationary scenario
`drives' the universe to a flat geometry \foreignlanguage{italian}{\cite{key-9,key-10}}.

\section{Conclusion remarks}

By using a formula that directly connects the average amplitude of
the relic SGWB with the Inflaton field and the equation for the characteristic
amplitude $h_{c}$ for the relic SGWB, in this paper the upper bounds
on the relic SGWB from the WMAP and LSC data have been translated
in lower bounds on the Inflaton field.

The results show that the value of the Inflaton field that arises
from the WMAP bound on the relic SGWB is totally in agreement with
the famous slow roll condition on Inflation \foreignlanguage{italian}{\cite{key-9,key-10}},
while the value of the Inflaton field that arises from the LSC bound
on the relic SGWB could not be in agreement with such a condition. 

Finally, we further emphasize the importance of the formula (\ref{eq: fi}).
If the GWs interferometers will detect the relic SGWB in next years,
such a formula will permit to directly compute the amount of Inflation
in the early Universe. Hence, the analysis in this paper has shown
that the detection of the SGWB will permit a direct measure of the
value of the Inflaton field by giving an extraordinary precious and
precise information on the early Universe's dynamics. In other words,
the detection of the SGWB will permit \emph{to auscultate} \emph{the
primordial gravity's breath}.

\end{document}